\documentclass[]{article}
\usepackage[margin=1in]{geometry}
\usepackage{times}
\usepackage{titlesec}
\usepackage{hyperref}
\usepackage{graphicx}
\usepackage{float}
\usepackage{amsmath}
\usepackage{booktabs}
\usepackage{caption}
\usepackage{fancyhdr}
\usepackage{multicol}
\usepackage{tcolorbox}
\usepackage{fancyvrb}
\usepackage{listings}
\usepackage{cleveref}
\usepackage{xcolor}
\usepackage{tikz}
\usetikzlibrary{arrows.meta, positioning, fit, calc, backgrounds}
\crefname{lstlisting}{Listing}{Listings}
\Crefname{lstlisting}{Listing}{Listings}

\lstdefinestyle{Python}{
    language=Python,
    basicstyle=\ttfamily\footnotesize,
    commentstyle=\color{gray},
    keywordstyle=\color{blue}\bfseries,
    stringstyle=\color{red},
    showstringspaces=false,
    tabsize=4,
    frame=single,
    captionpos=b,
    numbers=left,
    numberstyle=\tiny\color{gray},
    breaklines=true,
    title=\lstname
}

\definecolor{codebg}{RGB}{248,248,248}

\lstdefinestyle{Shell}{
    language=bash,
    basicstyle=\ttfamily\footnotesize,
    commentstyle=\color{gray},
    keywordstyle=\color{blue},
    stringstyle=\color{red!70!black},
    showstringspaces=false,
    frame=single,
    backgroundcolor=\color{codebg},
    breaklines=true,
    captionpos=b,
    numbers=none,
}

\titleformat{\section}{\large\bfseries}{\thesection}{1em}{}
\titleformat{\subsection}{\normalsize\bfseries}{\thesubsection}{1em}{}

\hypersetup{
    pdftitle={Mapping CVEs to MITRE ATT\&CK Techniques: A Curated Gold-Set
      Classifier and the Limits of LLM-Assisted Label Expansion},
    pdfauthor={Cédric Bonhomme and Alexandre Dulaunoy},
    pdfsubject={Pre-preprint submitted to hack.lu 2026},
    pdfkeywords={CVE, MITRE ATT\&CK, vulnerability mapping, multi-label
      classification, large language models},
    colorlinks=true, linkcolor=blue, citecolor=blue, urlcolor=blue
}

\title{\textbf{Mapping CVEs to MITRE ATT\&CK Techniques:\\
A Curated Gold-Set Classifier and the Limits of\\
LLM-Assisted Label Expansion}}
\author{
    Cédric Bonhomme \\
    \textit{Computer Incident Response Center Luxembourg (CIRCL)} \\
    \href{mailto:cedric.bonhomme@circl.lu}{cedric.bonhomme@circl.lu} [\href{https://openpgp.circl.lu/pks/lookup?op=get\&search=0xA1CB94DE57B7A70D}{57B7 A70D}]
    \and
    Alexandre Dulaunoy \\
    \textit{Computer Incident Response Center Luxembourg (CIRCL)} \\
    \href{mailto:alexandre.dulaunoy@circl.lu}{alexandre.dulaunoy@circl.lu}
}
\date{2026-07-23}

\begin{document}

\maketitle

\begin{center}
  \small\textbf{Pre-preprint notice.} This manuscript accompanies a talk and
  paper submitted for consideration at hack.lu 2026 (October 2026).
\end{center}


\begin{abstract}
We present a reproducible pipeline for mapping Common Vulnerabilities and
Exposures (CVEs) to MITRE ATT\&CK (Enterprise) techniques from free-text
vulnerability descriptions. Rather than rely on the
CWE$\rightarrow$\allowbreak CAPEC$\rightarrow$\allowbreak ATT\&CK
derivation chain --- whose
table-expansion artifacts we quantify --- we train a multi-label classifier
on a curated \emph{gold} set of $1{,}207$ CVEs assembled from expert MITRE
Center for Threat-Informed Defense (CTID) mappings; it roughly doubles the
recall@5 of a zero-shot embedding-similarity baseline and improves every
ranking metric. We then ask whether LLM-assisted labeling can extend the gold
set, and reach the answer only after three successively more rigorous
experiments, each of which overturned the last. A single-run comparison says
expansion \emph{degrades} the classifier; averaged over five random seeds the
effect appears to reverse into a small, consistent ranking gain; but an
independent replication and an expansion-size scaling sweep ($100$--$984$
added CVEs, five seeds per size) show the apparent gain was an evaluation
artifact: LLM labels at $\approx$0.39 agreement with the experts produce no
reliable improvement at any size, and at $\sim$1{,}000 added CVEs they
measurably degrade rare-technique coverage (macro-F1 $-0.04$). The mechanism
is evaluation noise, not the labels alone: selecting the best training
checkpoint on a small test split turns every reported metric into a maximum
over dozens of noisy evaluations --- enough for identical fixed-seed runs to
differ by $0.05$ in recall@5, and for a five-seed comparison to pass its own
consistency criterion and still be wrong. Under a corrected protocol
(validation-split checkpoint selection, released as the default) the
gold-only results hold (recall@5 $0.673 \pm 0.019$) and a re-run of the
decisive contrast confirms the null expansion verdict. A gold-size scaling
curve completes the picture: every metric improves monotonically with more
curated rows while LLM-labeled rows add nothing --- the classifier is
label-quality bound, not data bound. All datasets, models, code, and the
complete trainer logs behind every table are released.
\end{abstract}

\section{Introduction}
\label{sec:intro}

A Common Vulnerabilities and Exposures (CVE) record~\cite{mitre_cve} describes
\emph{what} is broken in a piece of software. Defenders, however, must reason
about \emph{how} such a flaw would be used by an adversary: which techniques it
enables, and what impact its exploitation would have. The MITRE ATT\&CK
knowledge base~\cite{strom2018attack} is the community's shared vocabulary for
that behavioral layer, and connecting a CVE to the ATT\&CK techniques it enables
lets a defender pivot directly from a vulnerability feed to detection coverage,
threat-informed prioritisation, and risk assessment. Today that connection is
made largely by hand, and with hundreds of thousands of published CVEs it
cannot be made manually at scale.

The mapping is also genuinely difficult to automate. A single CVE legitimately
implies several techniques, so the task is multi-label; the distribution of
techniques over CVEs is long-tailed; and the mapping is partly subjective,
because a CVE characterises a flaw while a technique characterises attacker
behavior around it, so even expert analysts disagree. A tempting shortcut is to
\emph{derive} techniques deterministically through the existing cross-framework
links --- CWE~\cite{mitre_cwe} to CAPEC~\cite{mitre_capec} to ATT\&CK --- but, as
we quantify in \cref{sec:cve2capec}, that chain is dominated by table-expansion
artifacts and is unsuitable as a training target.

We take a supervised approach anchored in trustworthy labels. We assemble a
small \emph{gold} set of CVE-to-ATT\&CK mappings from the expert curation of the
MITRE Center for Threat-Informed Defense (CTID)~\cite{ctid_attack_to_cve,
ctid_mappings_explorer}, train a multi-label classifier on it, and benchmark
that classifier against a training-free semantic-similarity baseline in the
style of prior work~\cite{abdeen2023smet}. We then confront the gold set's chief
weakness --- its size --- by asking whether a capable large language model (LLM)
can label additional CVEs well enough to extend it, under a strict
validate-before-trust protocol that first measures the LLM's agreement with the
gold analysts.

This paper makes four contributions:
\begin{enumerate}
  \item A published, provenance-tiered CVE$\rightarrow$ATT\&CK dataset that keeps
        expert \emph{gold} labels strictly separate from weak automatically
        \emph{derived} labels, so the two are never conflated in training.
  \item A supervised multi-label classifier that roughly doubles the recall@5
        of a zero-shot similarity baseline and improves every ranking metric,
        evaluated under an honest, analyst-oriented protocol.
  \item A quantified analysis of the noise in the
        CWE$\rightarrow$CAPEC$\rightarrow$ATT\&CK derivation chain, explaining
        why its labels cannot be trained on.
  \item A multi-seed, multi-size LLM label-expansion experiment showing that
        labels at $\approx$0.39 agreement provide \emph{no} reliable gain at
        any expansion size tested ($100$--$984$ CVEs) and degrade
        rare-technique coverage at the largest size --- together with a
        methodological post-mortem: a single-seed run and a five-seed
        comparison each produced a different, wrong verdict, because
        best-checkpoint selection on a small test split amplifies run-to-run
        noise beyond the effect size. We identify the mechanism and release a
        corrected training protocol.
\end{enumerate}

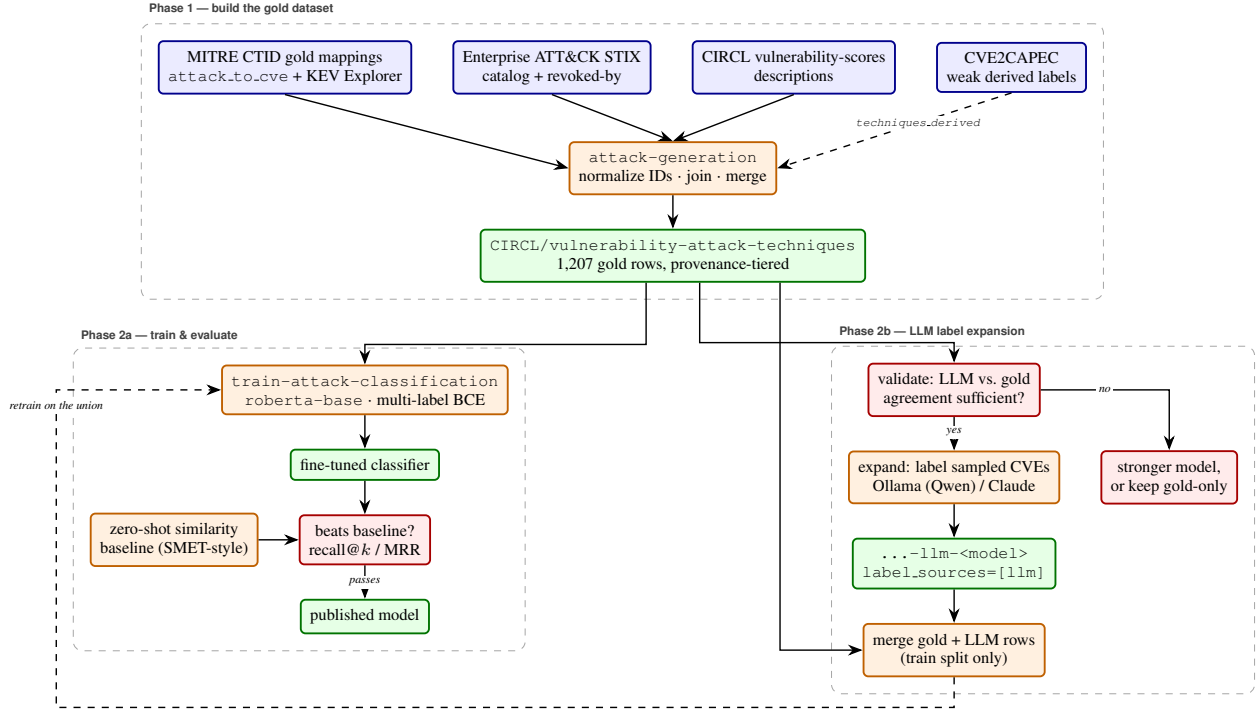
\begin{figure}[t]
\centering
\resizebox{\textwidth}{!}{%
\begin{tikzpicture}[
  font=\scriptsize,
  node distance=5mm and 6mm,
  every node/.style={align=center},
  data/.style={draw=blue!55!black, thick, fill=blue!8, rounded corners=2pt, inner sep=4pt},
  tool/.style={draw=orange!75!black, thick, fill=orange!12, rounded corners=2pt, inner sep=4pt},
  gate/.style={draw=red!65!black, thick, fill=red!8, rounded corners=2pt, inner sep=4pt},
  outp/.style={draw=green!45!black, thick, fill=green!10, rounded corners=2pt, inner sep=4pt},
  arr/.style={-{Stealth[length=2.2mm]}, semithick},
  lbl/.style={font=\tiny\itshape, fill=white, inner sep=1pt},
  phase/.style={draw=gray!70, dashed, rounded corners=4pt, inner sep=7pt},
  phaselbl/.style={font=\tiny\bfseries\sffamily, text=gray!50!black}
]
\node[data]                  (ctid)   {MITRE CTID gold mappings\\\texttt{attack\_to\_cve} + KEV Explorer};
\node[data, right=of ctid]   (stix)   {Enterprise ATT\&CK STIX\\catalog + revoked-by};
\node[data, right=of stix]   (scores) {CIRCL vulnerability-scores\\descriptions};
\node[data, right=of scores] (c2c)    {CVE2CAPEC\\weak derived labels};

\node[tool] (gen) at ($(stix.south)!0.5!(scores.south) + (0,-1.1)$)
  {\texttt{attack-generation}\\normalize IDs $\cdot$ join $\cdot$ merge};
\node[outp, below=of gen] (gold)
  {\texttt{CIRCL/vulnerability-attack-techniques}\\1{,}207 gold rows, provenance-tiered};

\draw[arr] (ctid.south)   -- (gen.west);
\draw[arr] (stix.south)   -- (gen.north);
\draw[arr] (scores.south) -- (gen.north);
\draw[arr, dashed] (c2c.south) -- (gen.east) node[lbl, pos=0.4] {\texttt{techniques\_derived}};
\draw[arr] (gen) -- (gold);

\node[tool] (train) at ($(gold.south) + (-4.6,-1.6)$)
  {\texttt{train-attack-classification}\\\texttt{roberta-base} $\cdot$ multi-label BCE};
\node[outp, below=of train] (model) {fine-tuned classifier};
\node[gate, below=of model] (eval)  {beats baseline?\\recall@$k$ / MRR};
\node[tool, left=of eval]   (base)  {zero-shot similarity\\baseline (SMET-style)};
\node[outp, below=of eval]  (done)  {published model};

\draw[arr] (train) -- (model);
\draw[arr] (model) -- (eval);
\draw[arr] (base)  -- (eval);
\draw[arr] (eval)  -- (done) node[lbl, pos=0.5] {passes};

\node[gate] (val) at ($(gold.south) + (4.2,-1.6)$)
  {validate: LLM vs.\ gold\\agreement sufficient?};
\node[tool, below=of val]   (exp)   {expand: label sampled CVEs\\Ollama (Qwen) / Claude};
\node[outp, below=of exp]   (llmds) {\texttt{...-llm-<model>}\\\texttt{label\_sources=[llm]}};
\node[tool, below=of llmds] (merge) {merge gold + LLM rows\\(train split only)};
\node[gate, right=of exp]   (stop)  {stronger model,\\or keep gold-only};

\draw[arr] (val)   -- (exp) node[lbl, pos=0.45] {yes};
\draw[arr] (val.east) -| (stop.north) node[lbl, pos=0.25] {no};
\draw[arr] (exp)   -- (llmds);
\draw[arr] (llmds) -- (merge);

\draw[arr] (gold.south) ++(-0.4,0) |- ($(train.north)+(0,0.3)$) -- (train.north);
\draw[arr] (gold.south) ++(0.4,0)  |- ($(val.north)+(0,0.3)$)   -- (val.north);
\draw[arr] (gold.south) ++(1.6,0)  |- (merge.west);
\draw[arr, dashed] (merge.south) -- ++(0,-0.45) -| ($(base.west)+(-0.5,0)$) |- (train.west)
  node[lbl, pos=0.45] {retrain on the union};

\begin{scope}[on background layer]
  \node[phase, fit=(ctid)(c2c)(gen)(gold)] (p1) {};
  \node[phase, fit=(train)(model)(base)(eval)(done)] (p2a) {};
  \node[phase, fit=(val)(exp)(llmds)(merge)(stop)] (p2b) {};
\end{scope}
\node[phaselbl, anchor=south west] at (p1.north west)  {Phase 1 --- build the gold dataset};
\node[phaselbl, anchor=south west] at (p2a.north west) {Phase 2a --- train \& evaluate};
\node[phaselbl, anchor=south west] at (p2b.north west) {Phase 2b --- LLM label expansion};
\end{tikzpicture}%
}
\caption{Overview of the pipeline. Blue nodes are external data sources,
orange nodes are VulnTrain commands, red nodes are decision gates, and green
nodes are published artifacts (Hugging Face). Phase 1 (\cref{sec:dataset})
builds the provenance-tiered gold dataset; Phase 2a
(\cref{sec:model,sec:eval}) trains the supervised classifier and
gates it on the zero-shot baseline; Phase 2b (\cref{sec:llm}) validates an
LLM labeler against the gold set before any expansion, and retraining on the
merged union is the experiment whose three-verdict history the paper
dissects (final verdict: keep gold-only).}
\label{fig:pipeline}
\end{figure}

All code (the VulnTrain toolkit), datasets, and models are released openly under
the CIRCL organisation so that every result --- including every stage of the
three-verdict expansion experiment --- can be reproduced. \Cref{fig:pipeline}
gives an overview of the full pipeline.

\section{Background and Related Work}
\label{sec:background}

\subsection{MITRE ATT\&CK}

ATT\&CK~\cite{strom2018attack} organises adversary behavior into tactics (the
\emph{why} of an action) and techniques (the \emph{how}), with many techniques
further refined into sub-techniques. It is a living knowledge base: techniques
are periodically added, deprecated, or revoked and replaced across versions. Any
system that consumes ATT\&CK identifiers over time must therefore normalise them
to a single reference version, which we do using the \texttt{revoked-by}
relationships published in the ATT\&CK STIX data~\cite{mitre_attack_stix}. We
target the Enterprise domain, whose version 19.1 catalogue contains $697$
active techniques and sub-techniques.

\subsection{The CTID ``Mapping ATT\&CK to CVE for Impact'' methodology}

Our labels follow the methodology defined by the MITRE Center for
Threat-Informed Defense~\cite{ctid_attack_to_cve}, which decomposes a CVE-to-
ATT\&CK mapping into three slots: the \emph{exploitation technique} used to
trigger the vulnerability, the \emph{primary impact} that directly results, and
any \emph{secondary impact} the primary one enables. We draw gold labels from
two CTID artifacts following this scheme: the 2021 \texttt{attack\_to\_cve}
mapping and the Known-Exploited-Vulnerabilities mappings distributed through the
CTID Mappings Explorer~\cite{ctid_mappings_explorer}, the latter keyed to the
CISA KEV catalogue~\cite{cisa_kev}. Together they form the gold set of
\cref{sec:dataset}.

\subsection{Automated CVE-to-ATT\&CK mapping}

Prior automated approaches fall into two broad families. The first
\emph{derives} techniques through the deterministic cross-framework chain
CWE$\rightarrow$CAPEC$\rightarrow$ATT\&CK. CVE2CAPEC~\cite{cve2capec} is a
widely used open implementation with near-complete coverage, and
BRON~\cite{hemberg2020bron} links the same threat and weakness catalogues into a
graph; both inherit the noise of the underlying mapping tables, which we measure
directly in \cref{sec:cve2capec}. The second family learns the mapping from
text. SMET~\cite{abdeen2023smet} deliberately avoids supervised classification
because of label scarcity, instead ranking techniques by semantic similarity
between a CVE description and ATT\&CK technique descriptions; this is precisely
the zero-shot baseline we adopt in \cref{sec:eval}. Closest to our supervised
setting, CVE2ATT\&CK~\cite{grigorescu2022cve2attack} fine-tunes a BERT model on a
few thousand CVEs over a restricted technique set. We differ in three ways: we
insist on expert CTID labels rather than derived ones, we adopt a multi-label
ranking evaluation aimed at analyst review rather than single-label accuracy,
and we report an explicit study of whether LLM-generated labels can extend the
gold set. A related CTID tool, TRAM~\cite{ctid_tram}, maps prose threat reports
to ATT\&CK, a different input distribution from terse CVE descriptions.

\subsection{LLMs as annotators}

Using an LLM to produce training labels in place of human annotators is
increasingly common, and structured-output decoding makes it practical to
extract schema-conformant technique assignments directly. The open question ---
which our expansion experiment (\cref{sec:llm}) addresses head-on for this task
--- is whether labels produced this cheaply are accurate enough to \emph{improve}
a supervised model, or whether their noise outweighs the extra coverage they
provide.

\section{Dataset Construction}
\label{sec:dataset}

The quality of the labels sets a hard ceiling on any supervised model, so the
central design decision of this work is where the training targets come from.
Two candidate sources exist: expert-curated mappings from the MITRE Center for
Threat-Informed Defense (CTID), and automatically \emph{derived} mappings from
the CWE\,$\rightarrow$\,CAPEC\,$\rightarrow$\,ATT\&CK cross-framework chain. We
evaluated both and, on the evidence below, train exclusively on the former
while retaining the latter as a clearly separated weak signal.

\subsection{Gold set from CTID mappings}

The gold set is assembled from two CTID sources that follow the ``Mapping
ATT\&CK to CVE for Impact'' methodology, in which each CVE is annotated with up
to three slots: the \emph{exploitation technique} an adversary would use to
trigger the flaw, the \emph{primary impact} that immediately follows, and any
\emph{secondary impact} that the primary one enables. The first source is the
2021 \texttt{attack\_to\_cve} mapping ($\sim$840 CVEs); the second is the
Mappings Explorer Known-Exploited-Vulnerabilities (KEV) set ($\sim$420 CVEs).
We fetch and join both, attach the corresponding CVE descriptions from the
CIRCL vulnerability-scores service, and reconcile them onto a single schema.

We stress that these labels are \emph{human} ground truth: both sources were
mapped manually by MITRE analysts applying the CTID methodology, not produced by
any automated tool. We do not annotate any CVE ourselves --- our contribution at
this stage is purely to fetch, normalise, merge, and split existing expert
mappings. This is precisely why we trust the gold set as a training target and
treat it as the reference against which every other label source (the
automatically derived labels of \cref{sec:cve2capec} and the LLM labels of
\cref{sec:llm}) is measured.

Because the two sources were authored against different ATT\&CK releases, the
same technique can appear under different identifiers across them (for example
T1562, revoked and replaced in a later version, versus its successor). Left
unnormalised, one behavioral concept would split into two distinct labels: the
classifier would see its training examples divided across both, inflating and
fragmenting the label vocabulary, and at evaluation a prediction of the new
identifier against a gold label carrying the old one would be scored as wrong
despite being correct. We therefore rewrite every identifier to a single
reference version by following the \texttt{revoked-by} relationships in the
ATT\&CK STIX data. The per-slot labels are then unioned into a flat
\texttt{techniques} field, which is the multi-label target the classifier
consumes. The final gold
set contains $1{,}207$ CVEs, partitioned into fixed train ($1{,}086$ CVEs) and
test ($121$ CVEs) splits that are reused unchanged throughout the paper; it is
published, with a DOI, as
\texttt{CIRCL/vulnerability-attack-techniques}~\cite{circl_gold_dataset}.

\subsection{Provenance tiers}

Every row records the origin of its labels in a \texttt{label\_sources} field
and keeps the weak derived labels, when present, in a separate
\texttt{techniques\_derived} column. This separation is a deliberate contract:
the derived column is \emph{never} used as a training target and never mixed
into the gold \texttt{techniques} field. Keeping it available nonetheless lets
downstream consumers filter cleanly back to gold-only data, and lets us study
where the deterministic chain diverges from analyst judgment --- which is
exactly the analysis of the next subsection.

\subsection{Quantifying CVE2CAPEC derivation noise}
\label{sec:cve2capec}

The derived labels come from CVE2CAPEC~\cite{cve2capec}, a daily updated
database that chains each CVE to techniques through the official
CWE\,$\rightarrow$\,CAPEC\,$\rightarrow$\,ATT\&CK cross-framework mappings. It is
an impressive piece of automation with near-complete coverage, and it is the
source most readily reached for when one needs CVE-to-ATT\&CK labels at scale.
We therefore examined directly whether its output is usable as a training
target, measuring on its \texttt{CVE-2024} database file of $39{,}156$ CVEs.

\begin{figure}[H]
\centering
\includegraphics[width=\textwidth]{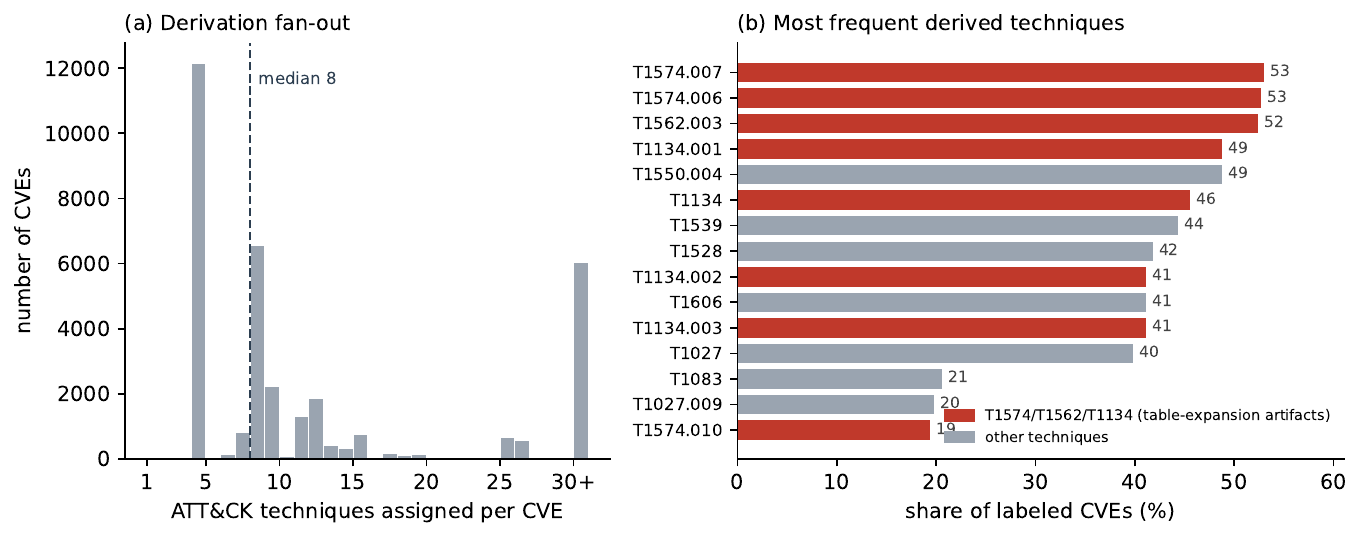}
\caption{Noise in the CVE2CAPEC CWE$\rightarrow$CAPEC$\rightarrow$ATT\&CK
derivation, measured on its \texttt{CVE-2024} database ($39{,}156$ CVEs;
$34{,}562$ labeled). \textbf{(a)} Distribution of the number of techniques
assigned per labeled CVE (capped at $30$ for display): the derivation attaches a
median of $8$ techniques to a CVE, with a heavy tail. \textbf{(b)} The fifteen
most frequent derived techniques as a share of labeled CVEs; the highlighted
T1574/T1562/T1134 cluster --- table-expansion artifacts with no per-CVE
justification --- each appears on roughly half of all CVEs. Training on these
labels would fit the structure of the mapping tables rather than adversary
behavior.}
\label{fig:cve2capec}
\end{figure}

Coverage is indeed excellent: $88.3\%$ of CVEs receive at least one technique.
The problem is not coverage but \emph{fan-out}, visualised in
\cref{fig:cve2capec}. The median labeled CVE is assigned $8$ techniques, the
mean $13.3$, and $7{,}381$ CVEs ($19\%$ of all CVEs) receive twenty or more
(\cref{fig:cve2capec}a). More tellingly, the marginal frequency of individual
techniques bears no relation to how vulnerabilities are actually exploited
(\cref{fig:cve2capec}b). The single most common technique across the database is
T1574.007 (\emph{Path Interception by PATH Environment Variable}), attached to
$53\%$ of all labeled CVEs, followed by T1574.006 and T1562.003 at roughly the
same level and a cluster of T1134 (\emph{Access Token Manipulation})
sub-techniques near $45$--$49\%$. No plausible reading of the vulnerability
landscape has path-interception implicated in a majority of all disclosed CVEs;
these frequencies are artifacts of \emph{table expansion}, in which a single
generic CWE fans out into dozens of CAPECs and, transitively, dozens of
techniques, regardless of the specific CVE. The highlighted bars in
\cref{fig:cve2capec}b are exactly this handful of expansion artifacts, which
between them are stamped onto roughly half the corpus.

A concrete example makes the failure mode tangible. CVE-2024-21732, a
cross-site-scripting vulnerability (CWE-79), is mapped through $48$ CAPECs to
techniques including T1027 (\emph{Obfuscated Files or Information}) and
T1574.006/\allowbreak.007 (\emph{Hijack Execution Flow}) --- none of which has any
relationship to cross-site scripting or drive-by exploitation. The label set is
determined by the weakness \emph{class} the CVE happens to be filed under, not
by the described behavior of the individual vulnerability.

The conclusion is that training on these labels would teach a model the
structure of the mapping tables rather than adversary behavior, and would in
particular inject a handful of spurious techniques onto the majority of
examples. We therefore exclude the derived labels from training entirely. They
remain useful in three narrower roles, all of which the retained
\texttt{techniques\_derived} column supports: as a candidate prior at inference
time (restricting suggestions to techniques compatible with the CWE chain), as
a baseline that any trained model must beat, and as a comparison column for
studying chain-versus-analyst divergence. We note that other CVE-to-ATT\&CK
resources built on the same chain, such as BRON, inherit the same noise profile
and are unsuitable as targets for the same reason.

\section{Supervised Classifier}
\label{sec:model}

Assigning ATT\&CK techniques to a CVE is intrinsically a \emph{multi-label}
problem: a single vulnerability legitimately implies several techniques across
the exploitation and impact slots. We therefore fine-tune a transformer encoder
with a sigmoid classification head and a binary cross-entropy objective
(\texttt{BCEWithLogitsLoss}), in which each technique is an independent
0/1 decision, rather than the softmax cross-entropy of a single-label
classifier. The model input is the concatenation of the CVE title and
description; the target is the gold \texttt{techniques} field of
\cref{sec:dataset}, and the derived labels are ignored entirely.

\paragraph{Why a compact classifier?} A central motivation for training a
dedicated encoder is operational efficiency, not only predictive quality.
Once fine-tuned, the classifier assigns all technique scores in a single
forward pass; it does not need to run a large autoregressive model or generate
and parse a textual answer for every incoming CVE. This makes low-latency batch
enrichment possible while reducing repeated computation and its associated
energy use. It also makes local deployment practical on constrained
infrastructure: the released model can run on a CPU when latency requirements
are modest, or on a small GPU for higher throughput, without sending embargoed
or otherwise sensitive vulnerability text to an external service. The goal is
therefore a model that national CSIRTs, small security teams, and open-data
publishers can operate continuously, rather than a capability available only
to organisations with large accelerators. We record training energy with
CodeCarbon, but a controlled comparison of inference latency and energy across
CPU, small-GPU, and generative-LLM deployments remains future work; the present
claim is deployability, not a measured energy-saving ratio.

Two design choices address the long-tailed label distribution. First, because
the gold set is small and heavily imbalanced across techniques, the loss uses
per-label positive weights: a technique's positive term is up-weighted in
inverse proportion to its frequency (clipped to a maximum ratio so that the
rarest labels cannot dominate the gradient). Second, the label space is
regularised structurally. Sub-techniques are collapsed to their parent
technique at training time --- there is rarely enough evidence in a terse CVE
description, or enough gold examples, to justify a sub-technique distinction ---
and the vocabulary is further restricted to techniques attested by at least a
minimum number of training examples (five by default). On the gold set this
yields a working vocabulary of $57$ parent techniques.

We fine-tune \texttt{roberta-base}~\cite{liu2019roberta} for $40$ epochs with the
AdamW optimiser, a learning rate of $1\times10^{-5}$ on a linear schedule, and a
fixed random seed of $42$; the checkpoint that maximises macro-F1 on an
evaluation split is retained --- \emph{which} split that should be turns out
to matter greatly, and is analysed and corrected in \cref{sec:llm}. Training
is orchestrated with the Hugging Face
\texttt{Trainer} and instrumented with CodeCarbon~\cite{codecarbon} so that the
energy cost of each run is recorded alongside its metrics (we analyse GPU
efficiency on this training infrastructure separately~\cite{bonhomme2025gpu}).
The framework
accepts several base encoders; we report \texttt{roberta-base} as the primary
configuration --- an encoder choice that follows our earlier VLAI model, a
\texttt{roberta-base} fine-tune for automated vulnerability severity
classification built with the same VulnTrain tooling~\cite{bonhomme2025vlai}.

\section{Evaluation Protocol and Baselines}
\label{sec:eval}

\paragraph{Metrics.} We evaluate the model as what it is intended to be --- a
tool that \emph{suggests} candidate techniques for an analyst to confirm --- so
ranking quality matters more than a single hard threshold. We report
\texttt{recall@}$k$ for $k\in\{3,5,10\}$, the fraction of a CVE's gold
techniques that appear in the model's top-$k$ ranked suggestions, together with
the mean reciprocal rank (MRR). We additionally report micro- and macro-averaged
F1 at a $0.5$ threshold. The distinction between the two averages is
substantive here: micro-F1 is dominated by the common techniques, whereas
macro-F1 weights every technique equally and therefore acts as our proxy for
\emph{rare}-technique performance --- the dimension the label-expansion
experiment of \cref{sec:llm} sets out to improve.

\paragraph{Zero-shot baseline.} As a reference point that requires no training
we implement a semantic-similarity ranker in the style of SMET: each ATT\&CK
technique is represented by the embedding of its official name and description,
each CVE by the embedding of its description (using a general-purpose sentence
encoder, all-MiniLM-L6-v2~\cite{reimers2019sbert, wang2020minilm}), and
techniques are ranked by cosine similarity to the CVE. This is a meaningful baseline because prior work deliberately turned to
similarity ranking precisely to avoid the label-scarcity that supervised
classification faces; if the fine-tuned model cannot beat it, the supervised
effort is not justified. Both systems are evaluated with the identical protocol
--- same test split, same label vocabulary, same metrics --- so the comparison
is direct.

\paragraph{Result.} \cref{tab:baseline} shows the fine-tuned classifier
improves on the zero-shot baseline across every ranking metric, roughly
doubling recall@3 and recall@5 (recall@5 rises from $0.322$ to $0.686$) and
lifting MRR from $0.397$ to $0.620$.\footnote{Single-run
numbers under the pre-correction evaluation protocol (see \cref{sec:llm});
under the corrected protocol the five-seed gold-only numbers of record are
recall@5 $0.673 \pm 0.019$ and macro-F1 $0.177 \pm 0.014$.} Even trained on
only $1{,}086$ gold examples ($1{,}083$ after the in-vocabulary filter),
supervision on trustworthy labels clearly pays off.
The remaining weakness is exactly where one would expect it --- rare-technique
performance, reflected in a macro-F1 of about $0.20$ on this run --- which
motivates the label-expansion study that follows.

\begin{table}[H]
\centering
\caption{Fine-tuned classifier (\texttt{roberta-base}, 57-technique vocabulary)
vs.\ the zero-shot similarity baseline, on the gold test split under an
identical protocol.}
\label{tab:baseline}
\begin{tabular}{lcccc}
\toprule
Method & recall@3 & recall@5 & recall@10 & MRR \\
\midrule
Zero-shot similarity (SMET-style) & 0.257 & 0.322 & 0.491 & 0.397 \\
Supervised (RoBERTa, gold)        & \textbf{0.482} & \textbf{0.686} & \textbf{0.842} & \textbf{0.620} \\
\bottomrule
\end{tabular}
\end{table}

\section{LLM-Assisted Label Expansion}
\label{sec:llm}

The gold set of \cref{sec:dataset} is trustworthy but small, and expert curation
does not scale to the millions of CVEs a production system must eventually
cover. A natural question is whether a large language model (LLM) can stand in
for the analyst and extend the gold set cheaply. This section describes a
labeler built for that purpose, a benchmark that selects the best available
model under a strict acceptance criterion, and a controlled experiment that
measures whether the resulting labels actually improve the classifier. The
answer, reached only after three rounds of increasingly rigorous experiments,
is negative: at $\approx$0.39 agreement the added labels produce no reliable
improvement at any expansion size we tested, and at the largest size they
degrade rare-technique coverage. The path to that answer --- a single run said
\emph{harmful}, five seeds said \emph{helpful}, and only an independent
replication plus a scaling sweep revealed \emph{no effect} --- is as much a
result as the verdict itself.

\subsection{Labeler design}

The labeler prompts an LLM to map a CVE description onto the same three-slot
schema the gold set uses --- exploitation technique, primary impact, and
secondary impact --- together with a one-sentence justification that records
\emph{why} each technique was assigned. Two interchangeable backends are
supported behind a single interface: a local model served by Ollama~\cite{ollama}, which
requires no API key and runs entirely on the user's own GPU, and the Anthropic
Claude API. All local-model labeling runs reported in this paper were served by
Ollama on a single server equipped with two NVIDIA H100 NVL GPUs; at that
capacity the 122B-parameter labeler processes a CVE in roughly 1.4 minutes, so
labeling one thousand CVEs is a day-long batch job. Both backends are
constrained to emit the answer as a structured object
matching the label schema, so parsing is not a source of error; technique
identifiers that fall outside the active ATT\&CK catalogue are dropped, and
malformed generations are retried.

Crucially, the labeler runs in one of two modes. In \texttt{validate} mode it
labels a held-out slice of the \emph{gold} set and reports the agreement between
the model's predictions and the analysts' labels, measured at the
parent-technique granularity the classifier is trained on. In \texttt{expand}
mode it labels \emph{new} CVEs that are not in the gold set, writing them to a
separate provenance tier (\texttt{label\_sources} $=$ \texttt{[llm]}) with the
backend, model identifier, and justification recorded on every row. The
validate-before-expand ordering is deliberate: expansion is only trustworthy if
the labeler first demonstrates, on data whose true labels are known, that it
agrees with the experts. This gate matters most for a local model, where the
alternative is to pay for a frontier API without knowing whether the cheaper
option is good enough.

We additionally exposed two levers intended to raise recall, which our early
runs identified as the weak point. The first is an \emph{assertive} prompt that
instructs the model to commit to a best exploitation technique and a best
primary impact for essentially every CVE, leaving a slot empty only when the
description genuinely supports nothing, while still forbidding it from padding a
slot with weakly related techniques. The second is an optional two-step
procedure (\texttt{--reason}): because a JSON-schema grammar forces a model to
emit its answer immediately, a \emph{thinking} model cannot reason before
committing, which tends to depress recall; the two-step mode therefore runs an
unconstrained analysis pass first and then a constrained extraction of the
technique identifiers from that analysis, at roughly double the per-CVE cost.

\subsection{Model-selection benchmark}

As a reference point for the labeler, we use the supervised classifier's
\emph{own} agreement with the gold labels. The intuition is that a labeler
agreeing with the experts less well than the trained classifier already does
might be expected to add little --- its labels are, on average, noisier than the
classifier's own predictions. On a matched full-split run the classifier reaches
a micro-F1 of $0.407$ against gold, which we treat as this reference level. We
emphasise \emph{reference} rather than hard gate: whether labels below this
level help or hurt in aggregate is an empirical question, which the expansion
experiment below answers directly, so we use the figure to contextualise the
labeler rather than to accept or reject it outright.

\cref{tab:llm-agreement} reports the benchmark. Three findings emerge. First,
\textbf{model capacity dominates prompt engineering}. On a common 30-CVE probe
the assertive prompt lifts the 35B-parameter model by only $0.02$ micro-F1,
whereas moving from the 35B to the 122B model lifts it by $0.13$ --- almost
entirely by repairing recall (from $0.27$ to $0.43$). The smaller model does
not disagree so much as \emph{under-commit}: it is reasonably precise when it
does answer but stays silent too often. Second, the two-step \texttt{--reason}
procedure \textbf{did not pay off} on the mid-size thinking model: on the same
30 CVEs it scored below the single-call configuration ($0.278$ versus $0.336$),
and its unconstrained reasoning pass repeatedly exceeded the inference timeout,
dropping whole CVEs to empty labels. It is not a substitute for capacity. Third,
and methodologically important, \textbf{small validation slices are
optimistic}: the 122B model scored $0.465$ on the 30-CVE probe but only $0.392$
on the full 121-CVE test split --- a $0.07$ drop, again concentrated in recall.
At $n=30$ the agreement estimate carries enough variance to mislead a go/no-go
decision, so we treat the full-split figure as the number of record and re-ran
the comparison rather than trusting the smaller sample.

The best configuration --- qwen3.5:122b, single call, assertive prompt --- thus
reaches $0.392$ micro-F1 against gold on the full split. This sits just
\emph{below} the classifier's $0.407$, but it is the strongest local
configuration available and its agreement is within the range commonly reported
for inter-analyst agreement on technique-level ATT\&CK CVE mappings. We
therefore treat it as the best-case candidate for expansion and test it
directly, rather than rejecting it on the benchmark alone.

\begin{table}[H]
\centering
\caption{LLM--vs--gold agreement (parent-technique level). Full split $= 121$
gold test CVEs.}
\label{tab:llm-agreement}
\begin{tabular}{llrrr}
\toprule
Model & Prompt / mode & P & R & micro-F1 \\
\midrule
qwen3.6:35b  & conservative, single & 0.429 & 0.248 & 0.314 \\
qwen3.6:35b  & assertive, single    & 0.442 & 0.271 & 0.336 \\
qwen3.6:35b  & assertive, --reason  & 0.395 & 0.214 & 0.278 \\
qwen3.5:122b & assertive, single (n=30)  & 0.509 & 0.429 & 0.465 \\
qwen3.5:122b & assertive, single (n=121) & 0.431 & 0.360 & \textbf{0.392} \\
\bottomrule
\end{tabular}
\end{table}

\subsection{Controlled expansion experiment}

To measure whether the selected labeler helps, we ran a deliberately small,
falsifiable pilot rather than committing to a large expansion up front. We
sampled 300 CVEs absent from the gold set and labeled them with the winning
configuration; 297 received at least one in-catalogue technique and were kept.
These rows were then folded into training and the classifier was retrained on
the resulting gold-plus-LLM union.

The experimental design isolates the effect of the LLM labels. The extra rows
are concatenated into the \emph{training} split only; the gold \emph{test}
split is left entirely untouched, so it remains a fixed yardstick against which
the union model and the gold-only model are directly comparable. Both models
were trained with identical code, random seed, and hyper-parameters --- the
297 LLM-labeled rows are the sole difference between the two runs. We also
re-measured the gold-only baseline under the current code rather than citing an
earlier figure, precisely because the model-selection benchmark had shown how
easily a stale or differently measured number can mislead.

The question is whether the LLM labels raise the model's quality on the gold
test split, and in particular whether they improve the analyst-facing ranking
metrics (\texttt{recall@}$k$) and the rare-technique proxy (macro-F1).

\paragraph{A single run gives the wrong answer.} Trained once, with the same
fixed seed used for the gold-only baseline, the union model appears
\emph{worse}: as \cref{tab:pilot} shows, every metric except micro-recall
regresses, with \texttt{recall@5} down $0.050$ and macro-F1 down $0.021$. Taken
at face value, this single comparison says expansion degrades the classifier,
and it is exactly the result our initial pilot reported. It is also unreliable,
though --- as two further rounds of experiments show --- not for the reason we
first diagnosed. We keep it here deliberately, because it is an instructive
cautionary example: in this small-data, high-variance regime a one-seed A/B
comparison is not merely imprecise, it can invert the sign of the effect.

\paragraph{Five seeds appear to reverse it.} We repeated both conditions across
five random seeds ($42$--$46$), which vary weight initialisation and data
shuffling, and report the mean $\pm$ standard deviation together with the paired
per-seed delta (\cref{tab:sweep}). The direction flips on the metrics that
matter: averaged over seeds the union model \emph{improves}
\texttt{recall@3} by $+0.038$, \texttt{recall@5} by $+0.030$ and micro-F1 by
$+0.020$, each formally consistent across seeds (paired
$|\Delta| > 2\,\mathrm{SEM}$), while macro-F1 stays flat. At the time we read
the single-seed pilot as the anomaly --- its gold-only \texttt{recall@5} of
$0.683$ sits roughly two standard deviations above the five-seed gold mean of
$0.641$ --- and adopted the ``small but consistent ranking gain'' as the
finding. The next two experiments overturned that reading: it was the five
gold-only runs of \cref{tab:sweep} that had drawn low.

\begin{table}[H]
\centering
\caption{The single-run pilot (seed $42$) on the gold test split; matched
training, LLM rows in \emph{train} only. Retained as a cautionary example: its
conclusion does not survive replication (cf.\ \cref{tab:sweep,tab:scaling}).}
\label{tab:pilot}
\begin{tabular}{lccc}
\toprule
Metric & Gold-only & Gold + LLM & $\Delta$ \\
\midrule
micro-F1     & 0.407 & 0.395 & $-0.012$ \\
macro-F1     & 0.185 & 0.164 & $-0.021$ \\
micro-recall & 0.625 & 0.626 & $+0.001$ \\
recall@3     & 0.546 & 0.491 & $-0.055$ \\
recall@5     & 0.683 & 0.633 & $-0.050$ \\
\bottomrule
\end{tabular}
\end{table}

\begin{table}[H]
\centering
\caption{Gold-only vs.\ gold+LLM over five seeds ($42$--$46$), \texttt{roberta-base}.
Mean $\pm$ std across seeds; $\Delta$ is the mean of the paired per-seed
differences. ``Consistent'' marks $|\Delta| > 2\,\mathrm{SEM}$. Retained as a
\emph{second} cautionary example: the apparent gain did not survive replication
(see text) --- its gold-only baseline had drawn low.}
\label{tab:sweep}
\begin{tabular}{lcccc}
\toprule
Metric & Gold-only & Gold + LLM & $\Delta$ (paired) & \\
\midrule
recall@3     & $0.506 \pm 0.019$ & $\mathbf{0.544 \pm 0.023}$ & $+0.038$ & consistent \\
recall@5     & $0.641 \pm 0.019$ & $\mathbf{0.670 \pm 0.033}$ & $+0.030$ & consistent \\
micro-F1     & $0.405 \pm 0.019$ & $\mathbf{0.424 \pm 0.010}$ & $+0.020$ & consistent \\
macro-F1     & $0.177 \pm 0.012$ & $0.173 \pm 0.017$ & $-0.004$ & within noise \\
micro-recall & $\mathbf{0.651 \pm 0.013}$ & $0.636 \pm 0.007$ & $-0.015$ & consistent \\
\bottomrule
\end{tabular}
\end{table}

\begin{figure}[H]
\centering
\includegraphics[width=\textwidth]{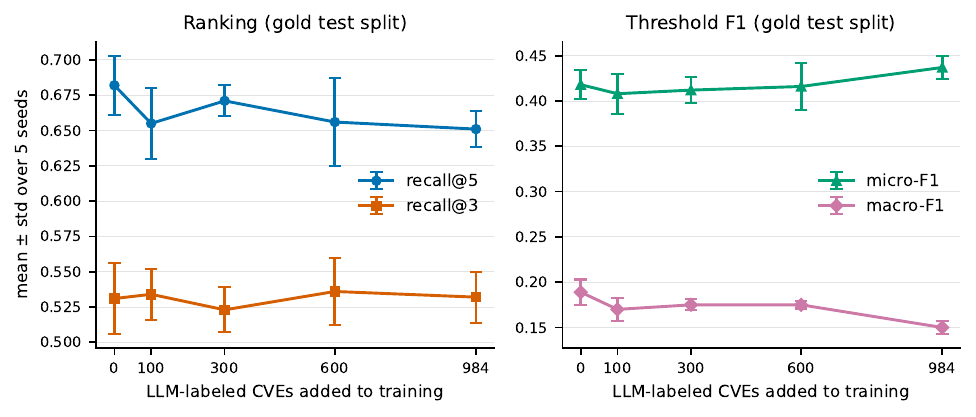}
\caption{The scaling sweep of \cref{tab:scaling} as a curve. \textbf{Left:}
ranking metrics on the gold test split as LLM-labeled CVEs are added to
training; \texttt{recall@5} drifts down, \texttt{recall@3} is flat.
\textbf{Right:} threshold metrics; micro-F1 edges up only at the largest size
while macro-F1 --- the rare-technique proxy that expansion was meant to
improve --- degrades. Error bars are $\pm$ one standard deviation over five
seeds.}
\label{fig:scaling}
\end{figure}

\paragraph{Replication and scale make the gain disappear.} Two further
experiments were designed to pin the effect down. First, a \emph{scaling
sweep}: we labeled an independent sample of $1{,}000$ new CVEs with the same
configuration ($984$ retained, published
as~\cite{circl_llm_scaling_dataset}) and retrained with the first
$N \in \{100, 300, 600, 984\}$ rows folded into training --- nested subsets,
so each size extends the previous one --- five seeds per size, against a
matched five-seed gold-only baseline from the same session
(\cref{tab:scaling}, \cref{fig:scaling}). No size reproduces the gain:
\texttt{recall@3} is flat everywhere, \texttt{recall@5} \emph{declines} with
size ($-0.031$ at $N{=}984$, $2.8\,\mathrm{SEM}$), and the only metric that
improves at the largest size, micro-F1 ($+0.019$, borderline at
$2.1\,\mathrm{SEM}$), does so while macro-F1 falls markedly ($-0.039$,
$5.6\,\mathrm{SEM}$ --- the largest and most consistent effect in the sweep):
the added labels densify the frequent techniques while drowning the rare ones.
Second, a direct \emph{replication}: we re-ran the original $297$-row union of
\cref{tab:sweep} in the same session as the new baseline. Its
\texttt{recall@5} reproduced almost exactly ($0.670 \pm 0.007$ against
$0.670 \pm 0.033$ before) --- but the matched gold-only baseline came out at
$0.682 \pm 0.021$, not $0.641$. The ``gain'' of \cref{tab:sweep} was never a
property of the union model; it was a low-drawn gold baseline.

\begin{table}[H]
\centering
\caption{Expansion-size scaling: gold training data plus the first $N$ rows of
an independently sampled, $984$-CVE LLM-labeled batch (nested subsets), five
seeds per size, evaluated on the untouched gold test split. Mean $\pm$ std.}
\label{tab:scaling}
\begin{tabular}{rcccc}
\toprule
$N$ & recall@3 & recall@5 & micro-F1 & macro-F1 \\
\midrule
$0$    & $0.531 \pm 0.025$ & $0.682 \pm 0.021$ & $0.418 \pm 0.016$ & $0.189 \pm 0.014$ \\
$100$  & $0.534 \pm 0.018$ & $0.655 \pm 0.025$ & $0.408 \pm 0.022$ & $0.170 \pm 0.013$ \\
$300$  & $0.523 \pm 0.016$ & $0.671 \pm 0.011$ & $0.412 \pm 0.014$ & $0.175 \pm 0.006$ \\
$600$  & $0.536 \pm 0.024$ & $0.656 \pm 0.031$ & $0.416 \pm 0.026$ & $0.175 \pm 0.004$ \\
$984$  & $0.532 \pm 0.018$ & $0.651 \pm 0.013$ & $0.437 \pm 0.013$ & $0.150 \pm 0.007$ \\
\bottomrule
\end{tabular}
\end{table}

\paragraph{The mechanism: evaluation noise from checkpoint selection.} What
made two internally consistent five-seed experiments disagree by $0.04$ on an
identical configuration? The training protocol selected, from $40$ per-epoch
checkpoints, the one with the best macro-F1 \emph{on the $119$-example test
split} (the trainer's in-vocabulary filter keeps $119$ of the $121$ gold test
CVEs), and reported that checkpoint's metrics. Late-training evaluations of so
small a split wobble substantially from epoch to epoch (\texttt{recall@5} spans
$0.62$--$0.70$ within a typical run), so the reported number is a
\emph{maximum over dozens of noisy evaluations} --- and ordinary GPU
nondeterminism (the order of floating-point reductions is not fixed by a fixed
seed) is enough to change which epoch wins. The direct evidence: three runs of
the identical gold-only configuration --- same seed, data, code, and hardware
--- reported \texttt{recall@5} of $0.636$, $0.683$, and $0.685$, a spread of
$0.048$ from evaluation noise alone. This defeats not only single-run
comparisons but also the paired $2\,\mathrm{SEM}$ criterion of
\cref{tab:sweep}, whose error model assumed per-run noise well below that
level. We have since corrected the protocol in the released code: checkpoint
selection now uses a validation split carved from the gold training data (the
test split is evaluated exactly once, for the final report), and an optional
fully deterministic mode makes fixed-seed runs bit-reproducible (single-GPU
only, at a substantial training-speed cost, since it serializes CUDA kernel
launches). The absolute
numbers in this paper predate the fix and therefore carry a mild
select-on-test optimism; every comparison, however, ran under the same
protocol on both sides. Re-running the gold-only configuration under the
corrected protocol (five seeds) re-establishes the numbers of record:
recall@5 $0.673 \pm 0.019$, recall@3 $0.536 \pm 0.032$, micro-F1
$0.410 \pm 0.006$, macro-F1 $0.177 \pm 0.014$ --- directionally about one
point below the pre-correction values and within noise of them, an upper
bound on the optimism that also absorbs the $10\%$ of training data ceded
to the validation split. Notably, the run-to-run standard deviation of
micro-F1 drops from $0.016$ to $0.006$: selecting checkpoints on a
dedicated split removes not only the bias but much of the variance.

\paragraph{The verdict under the corrected instrument.} As a final check, we
re-ran the decisive contrast under the corrected protocol: the same five
seeds, gold-only against the 297-row union and against all 984 LLM-labeled
rows (\cref{tab:contrast}). The conclusion is unchanged, and now rests on the
cleaner instrument rather than on post-hoc noise analysis. Neither expansion
configuration improves on gold-only ranking --- recall@5 drops from $0.673$
to $0.655$ (297 rows) and $0.651$ (984 rows) --- and at 297 rows \emph{every}
metric sits at or below the gold-only mean, leaving no trace of the
``consistent gain'' \cref{tab:sweep} once reported. The rare-technique
degradation at scale is confirmed at $\approx$2.9\,SEM (macro-F1
$0.177 \rightarrow 0.151$), and the borderline micro-F1 uptick persists
unresolved ($+0.017$, $\approx$1.3\,SEM).

\begin{table}[H]
\centering
\caption{The decisive contrast re-run under the corrected protocol
(validation-split checkpoint selection; five seeds, mean $\pm$ std). The
null verdict of \cref{tab:scaling} is confirmed: no ranking gain from LLM
rows at either size, and a confirmed macro-F1 degradation at 984.}
\label{tab:contrast}
\begin{tabular}{lcccc}
\toprule
Training data & recall@3 & recall@5 & micro-F1 & macro-F1 \\
\midrule
gold only            & $0.536 \pm 0.032$ & $\mathbf{0.673 \pm 0.019}$ & $0.410 \pm 0.006$ & $\mathbf{0.177 \pm 0.014}$ \\
gold $+$ 297 LLM rows & $0.511 \pm 0.023$ & $0.655 \pm 0.027$ & $0.404 \pm 0.012$ & $0.169 \pm 0.009$ \\
gold $+$ 984 LLM rows & $0.534 \pm 0.012$ & $0.651 \pm 0.022$ & $0.427 \pm 0.028$ & $0.151 \pm 0.014$ \\
\bottomrule
\end{tabular}
\end{table}

\begin{figure}[H]
\centering
\includegraphics[width=\textwidth]{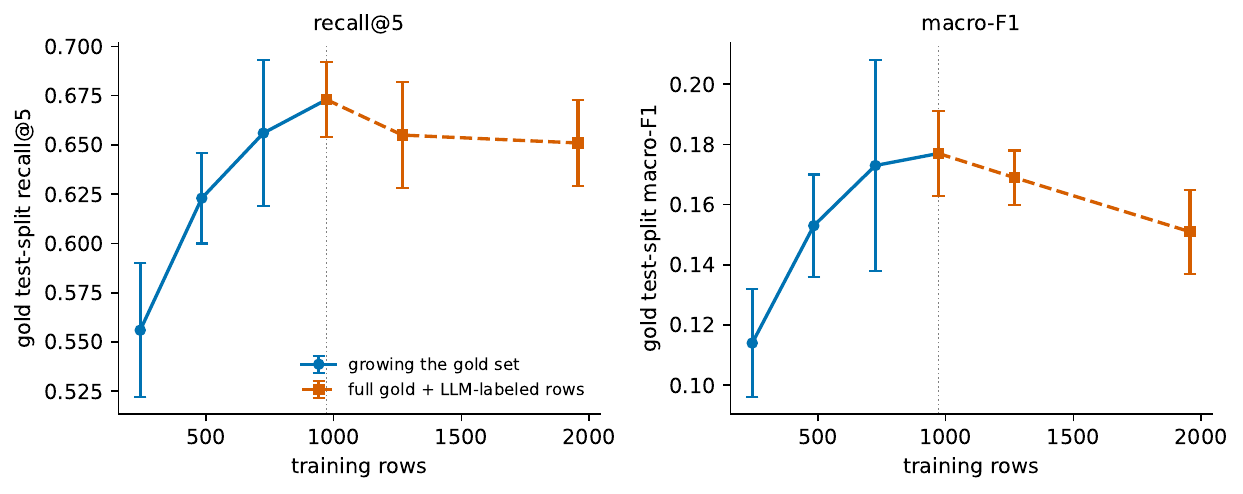}
\caption{Gold labels scale; LLM labels do not. Both panels share one x-axis
of training rows: the solid series grows the gold train split (nested
subsets, frozen label vocabulary and test set), the dashed series extends
the full gold set (dotted line, $972$ rows) with LLM-labeled rows under the
identical corrected protocol (\cref{tab:contrast}). Mean $\pm$ std over five
seeds. Recall@5 (left) and the rare-technique proxy macro-F1 (right) both
rise monotonically with curated data and flatten or degrade with
LLM-labeled data.}
\label{fig:goldcurve}
\end{figure}

\paragraph{Gold labels scale; LLM labels do not.} The natural control for
the null result is to grow the training set with labels we \emph{do} trust.
We trained on nested subsets of the gold train split ($25/50/75/100\%$, five
seeds each) under the corrected protocol, with the label vocabulary --- and
therefore the filtered test set --- frozen to the full-gold one so that only
training size varies.\footnote{The naive version of this experiment, with
the vocabulary rebuilt from each subset, \emph{inverts} the macro-F1 trend
(it reports $0.282$ at $243$ rows): smaller training sets produce smaller
label vocabularies, hence an easier averaging set and, through the
in-vocabulary test filter, an easier test set. One more instance of protocol
sensitivity in this regime; the released trainer freezes the vocabulary
under \texttt{-{}-train-fraction}.} Every metric rises monotonically with
gold size --- recall@5 from $0.556$ to $0.673$ and macro-F1 from $0.114$ to
$0.177$ between $243$ and $972$ rows.
\Cref{fig:goldcurve} overlays this curve with the expansion arm of
\cref{tab:contrast} on a single axis of training rows, and the contrast is
stark: the final $\sim$$250$ gold rows add $+0.017$ recall@5, while $297$
LLM-labeled rows added to the same full gold set \emph{subtract} $0.018$.
The classifier is not data-saturated --- gains are diminishing but still
positive at $972$ rows --- it is \emph{label-quality} bound: rows carrying
analyst-grade labels keep paying, rows at $0.39$ agreement do not, at any
tested size.

\paragraph{The mechanism, on one CVE.} Aggregate deltas of a few points can
feel abstract, so we close the experimental account with a single concrete
prediction, reproducible with the released single-CVE inference command
(\cref{lst:inference}). CVE-2021-44077 is an unauthenticated remote code
execution in Zoho ManageEngine ServiceDesk Plus, exploited in the wild; the
CTID analysts mapped it to twelve techniques, six of which survive
sub-technique collapse into the released vocabulary. It sits in the held-out
test split, so neither checkpoint saw it during training. Both released
checkpoints (seed $42$, corrected protocol;
gold-only~\cite{circl_released_model} and
gold$+$984~\cite{circl_expanded_model}) agree
on the head of the ranking --- exploit the public-facing application
(T1190), then a server software component, i.e.\ a web shell (T1505). But
the LLM-expanded checkpoint buys its extra confidence in the obvious call
(T1190: $0.72 \rightarrow 0.75$, rank $2 \rightarrow 1$) by deflating
nearly everything else the analysts credited to this CVE: T1505 drops below
the prediction threshold ($0.58 \rightarrow 0.45$), OS credential dumping
(T1003) falls from rank $8$ to $17$, account discovery (T1087) from $15$ to
$18$, and the tail technique T1136 (create account --- five training
examples) from rank $18$ to $32$. This is \cref{tab:contrast} in miniature
--- example-weighted confidence up, rare-technique ranking down --- though
a single CVE under a single seed pair is an illustration, not evidence; the
tables carry the evidence.

\begin{lstlisting}[style=Shell, label={lst:inference},
caption={One test-split CVE through both released checkpoints (output
abridged to the top five rows; \texttt{*}~=~predicted at the $0.5$
threshold, \texttt{+}~=~in the CTID gold mapping). The full ranking shows
the tail sinking: under the expanded checkpoint T1003 falls to rank 17 and
T1136 to rank 32.}]
$ vulntrain-infer-attack-classification --cve CVE-2021-44077 \
    --model CIRCL/vulnerability-attack-technique-classification-roberta-base

CVE-2021-44077 (test split)
gold: T1003 T1027 T1047 T1070 T1087 T1136 T1140 T1190 T1218 T1505 T1560 T1573
rank  technique   prob  pred  gold  name
   1  T1133       0.72   *          External Remote Services
   2  T1190       0.72   *     +    Exploit Public-Facing Application
   3  T1059       0.70   *          Command and Scripting Interpreter
   4  T1005       0.63   *          Data from Local System
   5  T1505       0.58   *     +    Server Software Component

$ vulntrain-infer-attack-classification --cve CVE-2021-44077 \
    --model CIRCL/vulnerability-attack-technique-classification-roberta-base-llm-expanded
            # gold + 984 LLM rows

rank  technique   prob  pred  gold  name
   1  T1190       0.75   *     +    Exploit Public-Facing Application
   2  T1059       0.71   *          Command and Scripting Interpreter
   3  T1133       0.68   *          External Remote Services
   4  T1210       0.50              Exploitation of Remote Services
   5  T1505       0.45         +    Server Software Component
\end{lstlisting}

\begin{figure}[H]
\centering
\includegraphics[width=\textwidth]{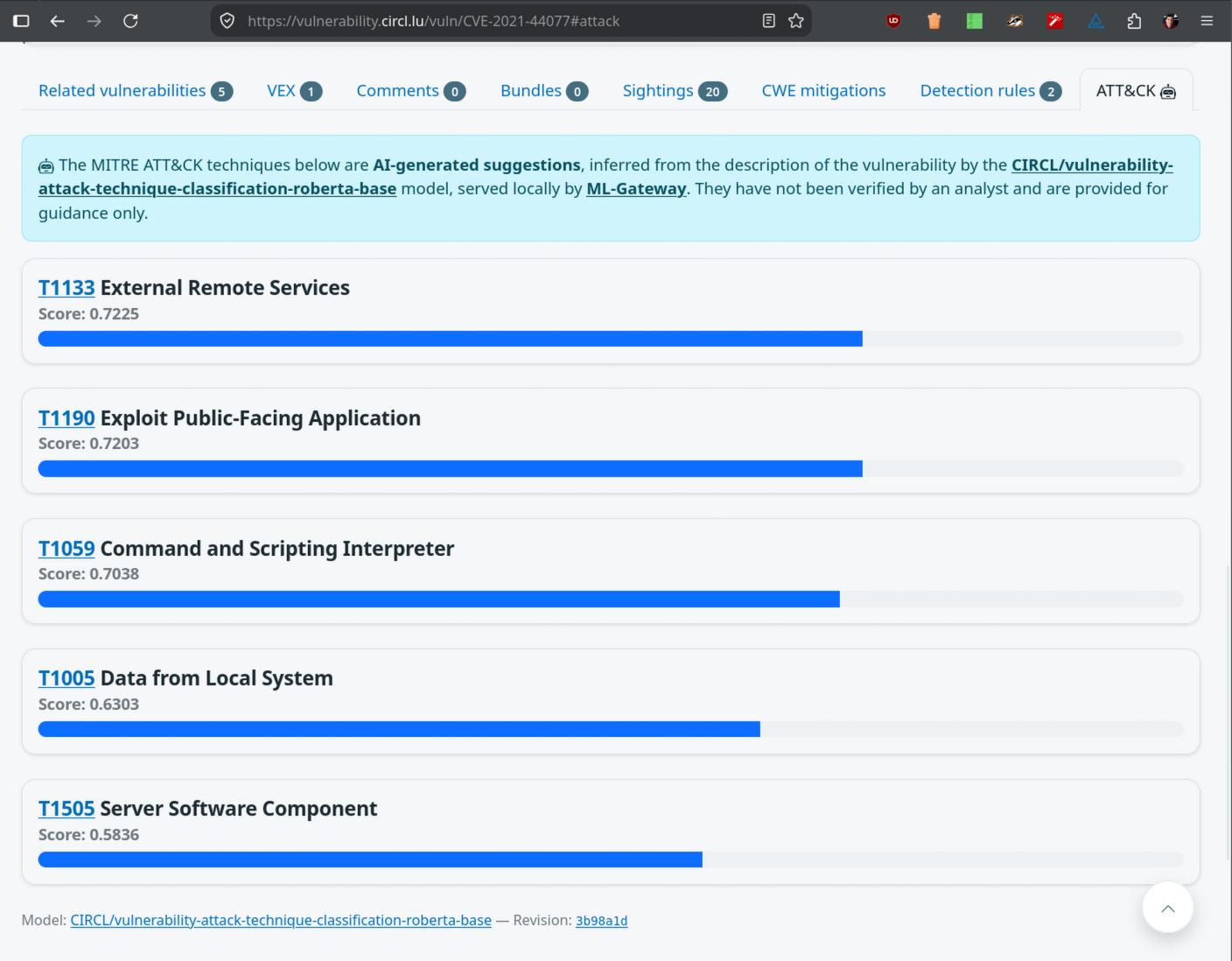}
\caption{The released gold-only checkpoint in production: the ATT\&CK tab of
CVE-2021-44077 on the public Vulnerability-Lookup instance
(\url{https://vulnerability.circl.lu}), where the model is served locally by
ML-Gateway. The five suggestions and their scores are those of
\cref{lst:inference}; the banner marks them as AI-generated suggestions not
verified by an analyst.}
\label{fig:deployment}
\end{figure}

\paragraph{In production.} The released gold-only checkpoint is not only a
research artifact: it is deployed on the public Vulnerability-Lookup
instance operated by CIRCL\footnote{\url{https://vulnerability.circl.lu}},
served locally by
ML-Gateway\footnote{\url{https://github.com/vulnerability-lookup/ML-Gateway}}.
Every vulnerability page carries an ATT\&CK tab with the model's ranked
suggestions, explicitly flagged as unverified AI-generated guidance.
\Cref{fig:deployment} shows that tab for the CVE of \cref{lst:inference} ---
the same five techniques and scores, as an analyst sees them in production.

\subsection{Interpretation}

Two conclusions follow, one about the task and one about method.

On the task: LLM labels at $\approx$0.39 agreement do not extend this gold
set. Across expansion sizes from $100$ to $984$ CVEs no ranking metric
reliably improves, and the clearest effect in the entire experiment is
negative: at the largest size the rare-technique proxy macro-F1 drops by
$0.039$. The pattern is what one would expect of labels whose errors
concentrate on exactly the techniques the gold set under-represents: frequent
techniques gain density (micro-F1 edges up at $N{=}984$) while the long tail
--- the very thing expansion was meant to fix --- is diluted by noise. The
result also retroactively vindicates the model-selection reference level
introduced above: a labeler that agrees with the experts slightly less well
than the trained classifier does ($0.392$ versus $0.407$) added nothing the
classifier did not already know. The gold scaling curve sharpens the
diagnosis: the classifier is not data-saturated --- every metric still rises
between $725$ and $972$ gold rows --- so the ceiling is the labels, not the
data volume, and enlarging the curated set is the one intervention with
measured payoff. Paths that could still work --- and that the
released machinery supports --- are higher-agreement labeling (a stronger
model, human-reviewed silver labels, or restricting to the labeler's
high-confidence slots) and targeting rare techniques explicitly rather than
sampling CVEs at random.

On method: the same underlying system produced three different verdicts ---
\emph{harmful} (one seed), \emph{helpful} (five seeds), \emph{no effect}
(replication and scaling) --- and the binding constraint turned out to be
neither label quality nor seed variance but the evaluation protocol itself.
Reporting variance across seeds, the standard prescription, was necessary but
not sufficient: the paired five-seed test passed its own consistency criterion
and was still wrong, because checkpoint selection on a small test split
injects per-run noise comparable to the effect size. Claims about label
expansion in this regime --- ours or others' --- need a selection split that
is not the test split, deterministic or repeated runs, and replication on an
independent sample before a small effect is believed. We retain the labeler,
both backends, the union-training path, and the corrected protocol in the
released code so the full comparison is reproducible.

\subsection{How the experiments were run}
\label{sec:howrun}

Every experimental result in this paper was produced the same way: a shell
loop over random seeds for each configuration, one self-contained trainer log
per run, and an aggregation script that extracts the single final test-split
evaluation from each log and reports mean~$\pm$~std per configuration. The
complete trainer logs are released alongside the code so that every table can
be regenerated from them. As an illustration, \cref{lst:contrast} shows the
corrected-protocol expansion contrast: the gold-only arm reuses the five runs
of record, and the expansion arm folds the LLM-labeled rows into the training
split via \texttt{-{}-extra-dataset-id}. The corrected protocol needs no
extra flags --- a gold-only validation split for checkpoint selection
(\texttt{-{}-val-split 0.1}) is the released default.

\begin{lstlisting}[style=Shell, label={lst:contrast},
caption={The corrected-protocol expansion contrast, as executed. The
297-row-union arm differs only in \texttt{-{}-extra-dataset-id}; the
expansion-size sweep of \cref{tab:scaling} additionally caps the folded-in
rows with \texttt{-{}-extra-max-rows}.}]
mkdir -p contrast-logs
# gold-only arm: reuse the five runs of record (seeds 42-46)
cp record-logs/size0-seed*.log contrast-logs/

# expansion arm: gold + all 984 LLM-labeled rows, five seeds
for s in 42 43 44 45 46; do
  vulntrain-train-attack-classification \
    --base-model roberta-base --seed $s --no-push \
    --repo-id tmp/contrast --model-save-dir results/contrast \
    --extra-dataset-id CIRCL/vulnerability-attack-techniques-llm-scaling \
    2>&1 | tee contrast-logs/size984-seed$s.log
done

# aggregate: final test-split metrics, mean +/- std per configuration
python3 scaling_sweep.py contrast-logs
\end{lstlisting}

\section{Discussion and Limitations}
\label{sec:discussion}

\paragraph{The right yardstick for absolute performance.} The absolute F1 scores
in this paper are low in machine-learning terms, and it would be a mistake to
read them against the near-ceiling accuracies typical of well-posed
classification. Mapping a CVE to ATT\&CK is genuinely hard and partly
subjective: a CVE description characterises a \emph{flaw}, while an ATT\&CK
technique describes \emph{attacker behavior} built around that flaw, and even
expert analysts disagree on the mapping. The agreement our best labeler reaches
with the gold analysts ($\approx$0.39 micro-F1) is, tellingly, in the same range
as the classifier's own agreement, which suggests both are pressing against an
inherent task ceiling rather than a modelling deficiency. The practical
consequence is that any model built on this data should be presented as
\emph{suggesting candidate techniques for analyst review}, not as an
authoritative mapping --- which is also why our headline metrics are the
top-$k$ ranking measures rather than a single hard-threshold score.

\paragraph{Gold-set size and selection bias.} The gold set is small
($\sim$1{,}200 CVEs) and not a representative sample of the CVE corpus. Both CTID
sources over-represent vulnerabilities exploited in the wild --- the KEV set by
construction --- so the technique distribution skews toward remote exploitation
of servers. Models trained here will inherit that skew, and reported numbers may
not transfer unchanged to the long tail of never-exploited CVEs. The size also
bounds the label vocabulary: techniques with too few examples are cut, so the
classifier simply cannot emit large parts of the ATT\&CK matrix.

\paragraph{Ground truth is scarce and difficult to validate.} Finding a usable
ground-truth dataset is itself a central difficulty of this task. Public
CVE-to-ATT\&CK collections are uncommon, often small, and sometimes generated
through the same CWE$\rightarrow$CAPEC$\rightarrow$ATT\&CK tables that a model
would subsequently be evaluated against. Such circular evaluation can reward a
model for reproducing table-expansion artifacts rather than for selecting the
techniques an analyst would apply to the individual CVE. Even expert mappings
are not automatically exhaustive: a terse CVE description may support several
plausible exploitation and impact paths, and the absence of a technique from a
record does not establish that it is a true negative. Ensuring that labels use
the proper technique therefore requires both identifier validation against a
specific ATT\&CK release and semantic review of whether each technique is
justified by the vulnerability. We use the CTID mappings as the best available
operational gold standard, but not as a claim that every valid technique for
every CVE has been enumerated.

\paragraph{Version and schema drift.} The 2021 CTID mappings were authored
against an older ATT\&CK release, while ATT\&CK is a changing ontology rather
than a fixed label list. Over time, techniques can be added, deprecated, revoked
and replaced; names and descriptions can change; and a former parent-level
concept can acquire sub-techniques that express distinctions unavailable to the
original annotator. Tactic associations and STIX relationships may change as
well, and consumers must also keep the Enterprise, Mobile, and ICS domains
separate. Our \texttt{revoked-by} normalisation resolves explicit identifier
replacements into the version used by this study, but it cannot infer a modern
sub-technique from an older coarse label, reconstruct a mapping after a semantic
split or merge, or decide whether revised technique guidance would change the
analyst's choice. Reproducible datasets should consequently record their ATT\&CK
release, preserve the original identifiers, and treat migration as a reviewed
data transformation rather than a blind string substitution. Restricting this
study to the Enterprise domain also discards a small number of
Mobile/ICS-tagged CVEs.

\paragraph{Scope of the expansion result.} Our expansion verdict --- no
reliable gain at $\approx$0.39 agreement, with a rare-technique cost at scale
--- is established for one local model family, one prompt configuration,
expansion sizes up to $984$ CVEs, and five seeds per configuration. What it
rules out is the hope that \emph{scale alone} rescues labels at this agreement
level: adding more of them, up to roughly the size of the gold set itself,
does not help and hurts the tail. It does not rule out expansion as such: a
frontier API model, a human-in-the-loop silver-labeling process, or a labeler
restricted to its high-confidence slots might clear an agreement bar that our
best local configuration does not, and the released machinery supports exactly
that re-run. A further threat to validity is that folding LLM rows into
training also enlarges the label vocabulary, so the union and gold-only models
are not graded on an identical label space; this is inherent to the treatment
(expanding coverage \emph{is} part of what expansion does) rather than a
confound we could remove, but it means the comparison answers the practical
question ``does adding these labels produce a better model?'' rather than a
fully label-space-controlled one. The absolute numbers of
\cref{sec:model,sec:eval} and of
\cref{tab:pilot,tab:sweep,tab:scaling} were
measured under the pre-correction protocol (checkpoint selection on the test
split) and carry a mild select-on-test optimism; comparisons are unaffected in
design, and the gold-only numbers of record have since been re-established
under the corrected protocol (validation-split checkpoint selection, five
seeds; \cref{sec:llm}), landing about one point lower and within noise of the
pre-correction values; the decisive expansion contrast was likewise re-run
under the corrected protocol and confirms the null verdict
(\cref{tab:contrast}). Finally, the
experiment measures classifier quality on the gold test split; it does not
directly measure label quality on the \emph{new} CVEs, which by definition have
no gold to check against.

\paragraph{Base-model choice.} We fine-tune \texttt{roberta-base} throughout
and did not search over base encoders. It is a strong, standard default for
text classification, but it is pretrained on general English and may
underweight the security-specific vocabulary of CVE text. The expansion result
itself is an \emph{internal} comparison in which the gold-only and
gold-plus-LLM models share the same encoder, so the \emph{direction} of the
effect --- no reliable gain from expansion at this agreement level, and a
rare-technique cost at scale --- is a within-encoder difference we would not
expect the base model to reverse. We tested that expectation directly with a
robustness check on ModernBERT-base~\cite{warner2024modernbert}: ten runs
under the identical corrected protocol (five seeds, gold-only and
gold$+$984; \cref{tab:basemodel}) rather than a re-run of the full
experimental grid. Both expectations held. The more recent encoder does not
raise the ceiling: gold-only micro- and macro-F1 are statistically
indistinguishable from \texttt{roberta-base} ($0.416 \pm 0.020$ vs.\
$0.410 \pm 0.006$; $0.180 \pm 0.014$ vs.\ $0.177 \pm 0.014$), while both
ranking metrics are clearly lower (recall@5 $0.614 \pm 0.018$ vs.\
$0.673 \pm 0.019$, $\approx$5\,SEM), so \texttt{roberta-base} remains the
released default. And the expansion verdict replicates: adding the $984$
LLM-labeled rows again degrades macro-F1 ($0.180 \rightarrow 0.152$,
$\approx$3.1\,SEM) and again yields no reliable recall@5 gain ($+0.021$,
$\approx$1.4\,SEM --- with the mean still below gold-only
\texttt{roberta-base}). The one quantitative difference is that the
borderline micro-F1 uptick of \cref{tab:contrast} is larger and resolves on
ModernBERT ($+0.034$, $\approx$2.7\,SEM), consistent with the mechanism
described above: LLM rows concentrated on head techniques can buy
example-weighted F1 precisely while eroding the tail. A systematic search
over encoders --- domain-adapted models such as SecureBERT, larger variants
such as \texttt{roberta-large} --- is left to future work; on this evidence
the bottleneck is label quality, not encoder recency.

\begin{table}[H]
\centering
\caption{Base-model robustness check: the corrected-protocol contrast
re-run on ModernBERT-base (five seeds, mean $\pm$ std;
\texttt{roberta-base} rows repeated from \cref{tab:contrast}). The
macro-F1 degradation from LLM rows replicates on the second encoder, and
ModernBERT does not improve on gold-only \texttt{roberta-base} ranking.}
\label{tab:basemodel}
\setlength{\tabcolsep}{5pt}
\begin{tabular}{llcccc}
\toprule
Encoder & Training data & recall@3 & recall@5 & micro-F1 & macro-F1 \\
\midrule
\texttt{roberta-base} & gold only            & $0.536 \pm 0.032$ & $0.673 \pm 0.019$ & $0.410 \pm 0.006$ & $0.177 \pm 0.014$ \\
\texttt{roberta-base} & gold $+$ 984 LLM rows & $0.534 \pm 0.012$ & $0.651 \pm 0.022$ & $0.427 \pm 0.028$ & $0.151 \pm 0.014$ \\
ModernBERT-base       & gold only            & $0.503 \pm 0.017$ & $0.614 \pm 0.018$ & $0.416 \pm 0.020$ & $0.180 \pm 0.014$ \\
ModernBERT-base       & gold $+$ 984 LLM rows & $0.524 \pm 0.023$ & $0.635 \pm 0.029$ & $0.450 \pm 0.020$ & $0.152 \pm 0.015$ \\
\bottomrule
\end{tabular}
\end{table}

\section{Conclusion}
\label{sec:conclusion}

We set out to map CVEs to MITRE ATT\&CK techniques from vulnerability
descriptions, and to do so honestly --- measuring against a real baseline and
subjecting our own conclusions to the scrutiny that small-data experiments
demand. Two results stand out. First, a supervised multi-label classifier
trained on a small, curated gold set of CTID mappings roughly doubles the
recall@5 of a zero-shot similarity baseline and improves every ranking
metric, confirming that
trustworthy labels, even in modest quantity, beat a training-free semantic
ranker for this task. Second, extending that gold set with a capable local LLM
agreeing with the experts at $\approx$0.39 does \emph{not} produce a better
classifier: across expansion sizes from $100$ to $984$ CVEs and five seeds per
size, ranking quality never reliably improves, and at the largest size
rare-technique coverage measurably degrades --- on both encoders we tested.
Cheap LLM labels at this agreement level are not, on this evidence, a
substitute for expert curation --- and in particular they are not the fix for
the long tail that motivated expansion in the first place.

The methodological result is at least as important as the empirical one. The
identical comparison produced three verdicts as rigor increased: one seed said
expansion is harmful, five seeds said it helps, and replication with an
independent sample plus a size sweep said it does nothing --- a verdict a
final re-run under the corrected protocol confirms. The culprit was not
seed variance alone but evaluation noise: selecting the best of $40$
checkpoints on a $119$-example test split makes every reported metric a
maximum over noisy evaluations, and identical fixed-seed runs differed by up
to $0.05$ in recall@5. Multi-seed reporting --- the standard prescription ---
was necessary but not sufficient; controlling the evaluation protocol (a
dedicated selection split, repeated or deterministic runs) is what finally
makes the answer stable.

\paragraph{Further work and an open-data path.} The immediate technical next
steps are higher-agreement labeling (a stronger model, human-reviewed silver
labels, or high-confidence slots only) aimed squarely at rare techniques;
CWE-stratified rather than random sampling of CVEs to label; evaluation on
newer and less exploitation-biased CVEs; and larger domain-adapted encoders.
The clearest measured opportunity, however, is to enlarge the curated gold
set: its scaling curve is still improving at the current size, whereas adding
unreviewed LLM labels is not. We therefore hope this release can seed a broader
open-data effort in which incident responders, vulnerability analysts,
researchers, and vendors contribute reviewed CVE-to-ATT\&CK mappings rather
than building further private, mutually incompatible collections. Such a
project should retain the evidence and rationale for each mapping, distinguish
exploitation from primary and secondary impact, record annotator agreement and
provenance, pin the ATT\&CK release while preserving original identifiers, and
publish reviewed corrections and versioned splits. Contributions targeted at
rare techniques and under-represented vulnerability classes would be
especially valuable. A growing, auditable community dataset would enable both
fairer benchmarks and periodic retraining, and is the most credible route we
see to improving this model beyond the ceiling demonstrated here. All code,
datasets, models, and trainer logs --- including every stage of the
three-verdict expansion experiment --- are released to provide a reproducible
starting point for that effort.

\paragraph{From source changes to vulnerability records.} A complementary
direction is to apply the classifier before a polished vulnerability
description exists. Security-fix commit messages can state the affected
component and failure mode, while an associated patch or diff --- when it is
available and safe to process --- can expose the concrete behavior that the
short prose of a CVE record omits. A multi-input variant could encode the
commit message, changed code, and available advisory text together, then
generate ranked ATT\&CK techniques with evidence linked back to the relevant
message or patch lines. Integrated with existing record-generation tooling,
this could support an end-to-end pipeline that creates draft vulnerability
information and technique mappings automatically for GCVE~\cite{gcve2026} and CVE publication
workflows, rather than adding ATT\&CK enrichment only after publication.
Achieving full automation would require a new, openly reviewable training set
linking fixes, patches, vulnerability records, and versioned ATT\&CK labels,
as well as evaluation for information leakage, repository and programming-
language bias, unavailable or incomplete patches, and identifier drift. Given
the consequences of publishing an incorrect authoritative record, the first
deployment should retain human approval and machine-readable provenance; full
automation should be enabled only after its precision is established on
prospective, previously unseen fixes.

\section*{Reproducibility and Artifacts}
\begin{itemize}
  \raggedright
  \item Paper and trainer logs: \href{https://github.com/vulnerability-lookup/cve-attack-mapping-paper}{github.com/vulnerability-lookup/cve-attack-mapping-paper}.
  \item Code: VulnTrain (\href{https://github.com/vulnerability-lookup/VulnTrain}{github.com/vulnerability-lookup/VulnTrain}).
  \item Dataset: \texttt{CIRCL/vulnerability-attack-techniques}~\cite{circl_gold_dataset},
        DOI \texttt{10.57967/hf/9621}.
  \item Expansion dataset: \texttt{...-attack-techniques-llm-scaling}~\cite{circl_llm_scaling_dataset}
        ($984$ LLM-labeled rows, the treatment of
        \cref{tab:contrast,tab:scaling}), DOI \texttt{10.57967/hf/9622}.
  \item Model:
        \texttt{CIRCL/vulnerability-attack-technique-classification-roberta-base}~\cite{circl_released_model}
        (gold-only, corrected protocol, seed $42$), DOI
        \texttt{10.57967/hf/9623}. The negative-result comparison
        checkpoint of \cref{lst:inference}
        is published alongside it as
        \texttt{...-roberta-base-llm-expanded}~\cite{circl_expanded_model},
        DOI \texttt{10.57967/hf/9624} --- for reproducibility, not for use.
\end{itemize}

\section*{Acknowledgements}

This work was carried out in the context of the
\href{https://www.science.nask.pl/en/research-areas/projects/12456}{AIPITCH}
project (AI-Powered Innovative Toolkit for Cybersecurity Hubs), which aims
to create advanced artificial-intelligence-based tools
supporting key operational services in cyber defence, including technologies
for early threat detection, automatic malware classification, and the
improvement of analytical processes through the integration of Large
Language Models. The project is led by the
\href{https://www.nask.pl}{NASK} National Research Institute (Poland); the
international consortium includes \href{https://www.circl.lu}{CIRCL}
(Computer Incident Response Center Luxembourg, Luxembourg),
\href{https://www.shadowserver.org}{The Shadowserver Foundation}
(Netherlands), NCBJ (National Centre for Nuclear Research, Poland), and ABI
Lab (Centre of Research and Innovation for Banks, Italy).

\section*{Funding}

This work was co-funded by \href{https://www.circl.lu}{CIRCL} and by the
European Union through the
\href{https://www.science.nask.pl/en/research-areas/projects/12456}{AIPITCH}
project. The European Union contribution was provided by the European
Cybersecurity Industrial, Technology and Research Competence Centre under
grant agreement No~101190545 (call
DIGITAL-ECCC-2024-DEPLOY-CYBER-06). Views and opinions expressed are
however those of the author(s) only and do not necessarily reflect those
of the European Union or the European Cybersecurity Industrial, Technology
and Research Competence Centre. Neither the European Union nor the
granting authority can be held responsible for them.

\bibliographystyle{plain}
\bibliography{references}

\end{document}